\documentclass[seceq]{ptptex}

\usepackage{wrapft}

\usepackage{graphicx}
\usepackage{comment}
\usepackage{amsmath,amssymb}
\usepackage{type1cm}
\usepackage{bm}

\newcommand\beq{\begin{equation}}
\newcommand\eeq{\end{equation}}
\newcommand\beqa{\begin{eqnarray}}
\newcommand\eeqa{\end{eqnarray}}

\newcommand\bk{{\bf k}}

\title{
Magnetic Susceptibility of Quark Matter}

\author{Koichi Sato and Toshitaka Tatsumi}

\inst{
Department of Physics, Kyoto University, Kyoto 606-8502, Japan
}

\abst{
Magnetic properties of quark matter is discussed by evaluating the spin
 susceptibility within Fermi-liquid theory. We take into account the dynamical and static screening effects.
 At finite temperature, an anomalous $T^2\ln T$ term for susceptibity 
is shown as a non-Fermi-liquid effect due to the dynamical screening of 
transverse gluons. 
}

\begin{document}

\maketitle

\section{Introduction}
Nowadays the phase diagram of QCD in density-temperature plane has been extensively studied 
theoretically and experimentally. 
In particular, matter at high density but not so high temperature should
be interesting, since the Fermi surface is a good concept there and
interesting correlations such as superconducting pairing in a clear
way. Such situation may be closely related to compact stars.
  
Here we study the magnetic properties of quark matter.
A possibility of spontaneous spin polarization has
been suggested by one of the authors \cite{tat00,nak03}, using the
one-gluon-exchange (OGE) interaction. 
If it is realized at moderate densities,
it may give a microscopic origin of strong magnetic field observed in
compact stars, especially magnetars.\cite{Har06} 
Here we evaluate the magnetic susceptibility of quark matter at zero and
finite temperature 
within Fermi-liquid theory~\cite{Bay04,bay76}, taking into account the 
screening effects for gluons.~\cite{sch99} Throughout this paper we try
to figure out the characteristic aspects of the magnetic properties,
inherent in gauge theories. One important aspect may be a
non-Fermi-liquid behavior. 

\section{Magnetic Susceptibility at $T=0$ within Fermi Liquid Theory}
In a recent paper we have studied the magnetic susceptibility at $T=0$ within 
the Fermi-liquid theory to figure out the screening effect for gluons on the
magnetic instability in quark matter \cite{tat082,tat08}.
The magnetic susceptibility is defined as
\beq
\chi_M=\sum_{f=u,d,s}\chi_M^f=\sum_{f=u,d,s}
\left.\frac{\partial\langle M\rangle_f}{\partial B}\right|_{B=0}
\eeq
with the magnetization $\langle M\rangle_f$ for each flavor, and its
divergence signals a magnetic instability for spontaneous spin
polarization, {\it ferromagnetism}.  
Magnetic susceptibility is then written in terms of the quasi-particle 
interaction,
\beqa
\chi_M=\left(\frac{g_D^F\mu_q}{2}\right)^2\frac{N(T)}{1+N(T)\bar
f^a}
\eeqa
where $g_D^F=g_D(|\bk|=k_F)$ is the
gyromagnetic ratio on the Fermi surface at $T=0$.
$N(T)$ is the average of the density of states near the Fermi surface
and $\bar f^a$ is 
the angle-average of the spin-dependent quasi-particle interaction on the Fermi surface.

At $T=0$, $N^{-1}(0)=(\pi^2/N_ck_F^2)v_F$ where $v_F$ is the Fermi
velocity written 
in terms of another Landau-Migdal parameter $f_1^s$. It is well-known
that there appear infrared (IR) singularities in the Landau-Migdal
parameters in gauge theories. To improve the IR behavior of the
quasi-particle interaction, we must take into account the screening
effects for the gauge field. The inclusion of the screening effects also
matches the argument of the hard-dense-loop (HDL) resummation \cite{sch99}. 

We have seen that the transverse mode only receives the dynamical
screening due to the Landau damping, so that it   
still gives logarithmic singularities for the Landau-Migdal parameters.
However, they cancel each other in susceptibility to give a finite result. 
On the other hand, the momentum of 
the longitudinal mode is effectively cut off by the Debye mass $m_D$ to
give the $g^4{\rm ln}(1/g^2)$ term in susceptibility. 
We also find that this term has an interesting behavior, depending on 
the number of flavors. 
\begin{wrapfigure}{l}{6.5cm}
\includegraphics[width=7cm]{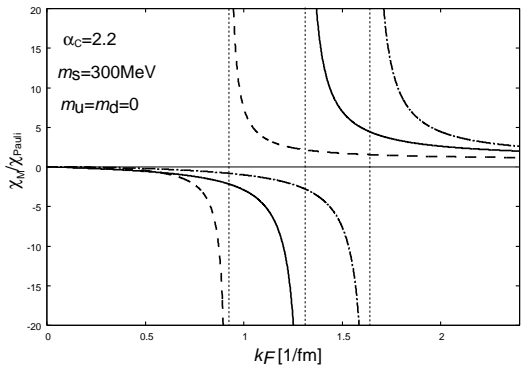}
\caption{Magnetic susceptibility at T=0. The solid curves shows the result using simple OGE, while the dashed and dash-dotted ones shows the screening effects with $N_f=1$(only $s$ quarks) and with $N_f=3$(u, d, and s quark), respectively.}
\end{wrapfigure}
In Fig. 1, we plot the magnetic susceptibility. \cite{tat08}
We assume a flavor-symmetric quark matter, and take the QCD coupling constant as
$\alpha_s=2.2$ and the strange quark mass $m_s=300$MeV inferred from the MIT bag model. 

One can see that the magnetic susceptibility diverges around
$k_F$ = 1-1.5 fm$^{-1}$ . The critical density for the
simple OGE without screening is consistent with the previous one
given by the energy calculation \cite{tat08}. One may expect that the quasi-particle
interaction is effectively cut off at momentum $|k|= m_D$,
which reduces the strength of the Fock exchange interaction, once
the screening is taken into account. However, this is not necessarily
the case in QCD. Compare the screening effects by changing
the number of flavors; if we change $N_f$, the screening effect
exhibits the opposite behavior for $N_f=1$ and
$N_f=3$ . In the case of $N_f= 1$ the screening effect 
works against the magnetic phase transition as in QED. 
Consequently the screening effect does not 
necessarily work against the magnetic instability, which is a different
aspect from electron gas.

\section{Magnetic susceptibility at finite temperature}
At finite temperature, we must evaluate $N(T)$, 
\begin{equation}
N(T)=-2N_c\int\frac{d^3k}{(2\pi)^3}\frac{\partial n_F(\epsilon_\bk)}{\partial\epsilon_\bk},
\end{equation}
with the Fermi-Dirac distribution function $n_F(\omega)$ and the
quasi-particle energy $\epsilon_\bk$.
The quasi-particle energy
is defined as $\epsilon_\bk=E_\bk+\Sigma_+(\epsilon_\bk,\bk)$, and the
leading order contribution of the self-energy $\Sigma_+(\omega,\bk)$ is
given by the one-loop diagram \cite{boy01},  
\begin{equation}
\Sigma_+(\omega,\bk)\sim -\frac{C_f g^2
 u_F}{12\pi^2}(\omega-\mu){\rm ln}\frac{\Lambda}{|\omega-\mu|}
\end{equation}
near the Fermi surface, 
where $u_F\equiv p_F/\mu$ and $C_f=(N_c^2-1)/2N_c$. 
The anomalous logarithmic term comes from the dynamical screening effect
for the transverse gluons, and gives rise to the non-Fermi-liquid
effects such as in the specific heat \cite{boy01,ipp,hol}. 
Taking into account the temperature dependence of the chemical
potential, we eventually find \cite{sat}
\begin{align}
\chi^{-1}_M  &= \chi_M^{-1}(T=0)
+\frac{\pi^4}{6N_ck_F^5E_F} \left(2E_F^2-m^2+\frac{m^4}{E_F^2} \right)\left(T^2
+\frac{C_fg^2u_F}{3\pi^2}T^2\ln\left(\frac{\Lambda}{T}\right) \right).
\label{chi_finT}
\end{align}
There appears $T^2 \ln T$ dependence on temperature because of the
anomalous behavior of the self-energy at the Fermi surface.  
This $T^2 \ln T$ dependence is a novel non-Fermi-liquid effect in
magnetic susceptibility, and corresponds to the $T \ln T$ dependence of the specific heat\cite{hol,ipp}.

Finally we show a magnetic phase diagram on the density-temperature
plane in Fig. 2, using the same parameter as in Fig.~1.
Compare the result with the full expression with the one without the dynamical screening effect {\it i.e.} $T^2 \ln T$ dependence. 
In the case without the $T^2 \ln T$ term, 
the critical temperature (Curie temperature) of the ferromagnetic phase
is about $T=60$MeV, while it can be at most $T=40$MeV including $T^2 \ln T$ dependence.
It turns out that the dynamical screening works against the magnetic
instability and can substantially shrink the ferromagnetic region on the phase diagram.
 

\begin{figure}
\begin{center}
\includegraphics[width=10cm]{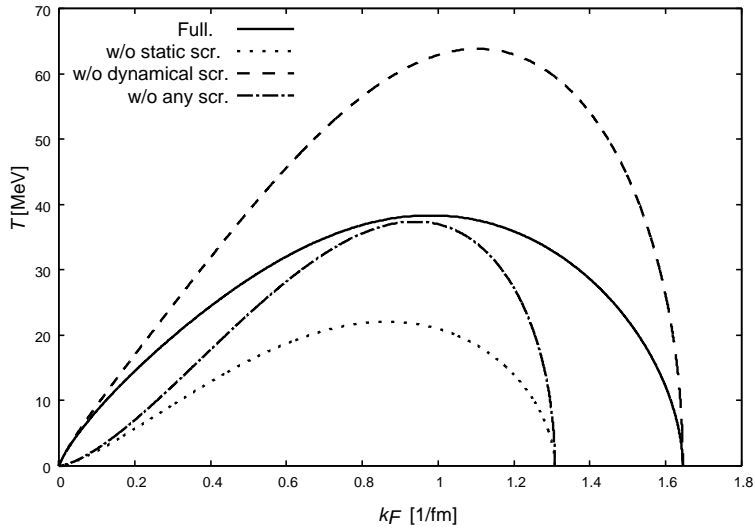}
\caption{Magnetic phase diagram on the density-temperature plane. The solid, dotted, dashed, and dash-dotted curves are the results for
the full expression, without the static screening, without the dynamical screening, and without any screening effect respectively.}
\end{center}
\end{figure}

 
The dash-dotted curve is the result without the static screening or $g^4
\ln g^{-2}$ term.~\cite{tat08} 
The static screening effect works in favor of the
ferromagnetic region, as discussed in the previous section. 

\section{Summary}
We have discussed some magnetic aspects of quark matter by evaluating
the magnetic susceptibility at both zero and finite temperature, based on
QCD, where the screening effects for gluons are taken into account.
At $T=0$, the static screening gives 
the term proportional to $m_D^2
\ln m_D^{-1}$ ; the Debye mass $m_D (\propto g^2)$ works as an infrared (IR) cutoff to 
remove the IR singularity in the exchange of the longitudinal gluons, while
it is still left in the quasi-particle
interaction due to the transverse gluons. 
At finite temperature, the dynamical screening effect gives the
anomalous term proportional to $T^2 \ln T^{-1}$ ;
the Fermi surface is smeared by $O(T)$ and the energy transfer of $O(T)$ is allowed among the
quasi-particles near the Fermi surface, so that temperature 
itself plays a role of the IR cutoff through the dynamical
screening effect for the transverse gluons.  
This logarithmic temperature dependence  
is a novel non-Fermi-liquid effect and its origin is the same as the well-known 
$T\ln T$ dependence of the specific heat.   
The anomalous $T^2 \ln T^{-1}$ term works against the magnetic
instability, as well as the usual $T^2$ term. 
We have seen that $T^2 \ln T^{-1}$ term should give rise to a sizable effect on
the magnetic phase diagram in the temperature-density plane.


\begin{thebibliography}{99}
\bibitem{tat00} T. Tatsumi, Phys. lett. {\bf B489} (2000) 280.\\
                T. Tatsumi, E. Nakano and K. Nawa, {\it Dark
	Matter}, p.39 (Nova Science Pub., New York, 2006).

\bibitem{nak03} E. Nakano, T. Maruyama and T. Tatsumi, Phys. Rev. {\bf
	D68} (2003) 105001.\\
                T. Tatsumi, E. Nakano and T. Maruyama,
	Prog. Theor. Phys. Suppl. {\bf 153} (2004) 190.\\
                T. Tatsumi, T. Maruyama and E. Nakano, {\it Superdense
	QCD Matter and Compact Stars}, p.241 (Springer, 2006).


\bibitem{Har06} A.K. Harding, D. Lai, Rep. Prog. Phys. 69 (2006) 2631.
\bibitem{Bay04}G. Baym, C.J. Pethick, Landau Fermi-Liquid Theory, Wiley-VCH, 2004;\\
P. Nozieres, Theory of Interacting Fermi Systems, Westview Press, 1997. 

\bibitem{bay76} G. Baym and S.A. Chin, Nucl. Phys. {\bf A262} (1976)
	527.
\bibitem{sch99} J.I. Kapusta, {\it Finite-temperature field
	theory} (Cambridge U. Press, 1993).\\
                M. Le Bellac, {\it Thermal Field Theory} (Cambridge
	U. Press, 1996).








\bibitem{tat08} T. Tatsumi and K. Sato, Phys. Lett. {\bf B663} (2008) 322.

\bibitem{tat082} T. Tatsumi, {\it Exotic States of Nuclear Matter} (World
	Sci., 2008) 272.

\bibitem{boy01} D. Boyanovsky and H.J. de Vega, Phys. Rev. {\bf D63} (2001) 034016.\\  C. Manuel, Phys. Rev. {\bf D55} (1997) 3215.

\bibitem{ipp} A. Ipp, A. Gerhold and A. Rebhan, Phys. Rev. {\bf D69}
	(2004) 011901.\\
A. Gerhold, A. Ipp and A. Rebhan, Phys. Rev, {\bf D70} (2004) 105015; PoS (JHW2005) 013.









\bibitem{hol} T. Holstein, R.E. Norton and P. Pincus, Phys. Rev. {\bf
	B8} (1973) 2649.












\bibitem{sat} K. Sato and T. Tatsumi, in preparation (2008).
\end{thebibliography}
\end{document}